\documentclass{article}
\usepackage{graphicx}
\usepackage{amsmath}
\usepackage{amssymb}
\usepackage{amsthm}
\usepackage{booktabs}
\usepackage{natbib}
\usepackage{hyperref}
\hypersetup{
  pdftitle={Uncertainty-Aware Deep Hedging},
  pdfauthor={Manan Poddar},
  colorlinks=true,
  linkcolor=blue,
  citecolor=blue,
  urlcolor=blue
}
\usepackage{tikz}
\usepackage[margin=1in]{geometry}
\usetikzlibrary{arrows.meta, positioning, fit, backgrounds, calc}
\title{Uncertainty-Aware Deep Hedging}
\author{Manan Poddar \\
\textit{Department of Mathematics} \\
\textit{London School of Economics and Political Science} \\
\texttt{m.poddar4@lse.ac.uk}}
\date{March 2026}

\begin{document}

\maketitle
\begin{abstract}
Deep hedging trains neural networks to manage derivative risk under market frictions, but produces hedge ratios with no measure of model confidence---a significant barrier to deployment. We introduce uncertainty quantification to the deep hedging framework by training a deep ensemble of five independent LSTM networks under Heston stochastic volatility with proportional transaction costs. The ensemble's disagreement at each time step provides a per-time-step confidence measure that is strongly predictive of hedging performance: the learned strategy outperforms the Black--Scholes delta on approximately 80\% of paths when model agreement is high, but on fewer than 20\% when disagreement is elevated. We propose a CVaR-optimised blending strategy that combines the ensemble's hedge with the classical Black--Scholes delta, weighted by the level of model uncertainty. The blend improves on the Black--Scholes delta by 35--80 basis points in CVaR across several Heston calibrations, and on the theoretically optimal Whalley--Wilmott strategy by 100--250 basis points, with all improvements statistically significant under paired bootstrap tests. The analysis reveals that ensemble uncertainty is driven primarily by option moneyness rather than volatility, and that the uncertainty--performance relationship inverts under weak leverage---findings with practical implications for the deployment of machine learning in hedging systems.
\end{abstract}
\section{Introduction}\label{sec:introduction}

The hedging of derivative securities is a cornerstone of modern financial risk management. In the classical Black--Scholes framework, a continuously rebalanced portfolio of the underlying asset can perfectly replicate the payoff of a European option, eliminating all risk. In practice, however, this idealisation breaks down: markets exhibit stochastic volatility, trading incurs transaction costs, and rebalancing can occur only at discrete intervals. Under these realistic conditions, perfect replication is impossible and the hedging problem becomes one of risk minimisation---choosing a trading strategy that keeps the residual profit and loss (P\&L) as small and predictable as possible.

Deep hedging, introduced by \citet{Buehler2019}, reframes this problem as a stochastic optimisation task solved by neural networks. Rather than deriving a hedging rule from a parametric model, a network is trained end-to-end on simulated price paths to minimise a chosen risk measure of the terminal P\&L, with transaction costs and other frictions incorporated directly into the objective. The approach has demonstrated strong results in simulation \citep{Buehler2019, Horvath2021} and is gaining traction in industry.

Yet a fundamental obstacle remains between the research literature and practical deployment. A trained deep hedging model outputs a single hedge ratio at each decision point---a point estimate with no accompanying measure of confidence. A risk manager presented with a recommendation to hold 0.47 units of the underlying has no way to assess whether the model is highly confident in this figure or whether it might equally well have recommended 0.30 or 0.65. In domains such as medical imaging and autonomous driving, this problem has been addressed through uncertainty quantification (UQ)---techniques that produce not just a prediction but a calibrated measure of the model's confidence in that prediction \citep{Gal2016, Lakshminarayanan2017}. In the deep hedging literature, however, UQ has received almost no attention.

This paper bridges that gap. We apply deep ensembles \citep{Lakshminarayanan2017}---a simple and effective UQ method in which multiple independently trained networks are queried simultaneously, with their disagreement serving as a measure of uncertainty---to the deep hedging framework. We then ask: can this uncertainty signal be used to improve hedging outcomes? Specifically, we propose an uncertainty-aware blending strategy that combines the ensemble's learned hedge with a classical Black--Scholes delta hedge, with the relative weighting determined by the ensemble's confidence at each time step. The blending parameters are optimised to minimise a tail-risk measure (CVaR), directly targeting the dimension on which neural hedging strategies are most vulnerable.

Working within a Heston stochastic volatility environment with proportional transaction costs, we find that ensemble uncertainty is a reliable predictor of hedging performance. When the ensemble members agree, the learned strategy outperforms Black–Scholes delta on approximately 80\% of simulated paths; when they disagree, it wins on fewer than 20\%. A CVaR-optimised blend that incorporates this signal achieves the best tail risk of any strategy tested, improving on Black--Scholes delta by 35--80 basis points in CVaR across three market calibrations, and on the theoretically optimal Whalley--Wilmott strategy by 100--250 basis points. These improvements are statistically significant and stable across calibrations.

Beyond the performance improvements, the analysis uncovers several findings of independent interest:
\begin{itemize}
    \item Ensemble uncertainty is driven primarily by moneyness rather than volatility: the models disagree most on deep in-the-money paths in calm markets, not during volatile episodes.
    \item The relationship between uncertainty and performance is regime-dependent, inverting completely when the stock--volatility correlation is weakened---a finding with implications for how UQ should be deployed across different market environments.
    \item The choice of risk measure fundamentally determines the optimal blending behaviour: an entropic risk criterion leads to near-complete reliance on the ensemble, while a CVaR criterion favours a classical-dominated blend. The same uncertainty signal supports both strategies.
    \item The ensemble's advantage over classical hedging arises from transaction cost savings (trading more selectively) rather than from superior price prediction---a nuance that informs how the blend should be constructed.
\end{itemize}

The remainder of this paper is organised as follows. Section~\ref{sec:related_work} reviews related work in deep hedging, uncertainty quantification, and classical hedging under transaction costs. Section~\ref{sec:methodology} describes the market model, network architecture, ensemble construction, blending mechanism, and evaluation methodology. Section~\ref{sec:results} presents the main empirical results. Section~\ref{sec:robustness} examines robustness across Heston calibrations, random seeds, and alternative specifications. Section~\ref{sec:discussion} discusses the economic interpretation of the findings, and Section~\ref{sec:conclusion} concludes.

\section{Related Work}\label{sec:related_work}

\paragraph{Deep hedging.} The deep hedging framework was introduced by \citet{Buehler2019}, who showed that feedforward neural networks trained to minimise convex risk measures can learn near-optimal hedging strategies in the presence of transaction costs, outperforming the classical Black--Scholes delta under Heston stochastic volatility. \citet{Horvath2021} extended this line of work to rough volatility models, demonstrating that recurrent architectures are necessary to capture the non-Markovian dynamics induced by fractional Brownian motion. \citet{Imaki2023} developed the \texttt{pfhedge} library, providing an open-source PyTorch implementation of the deep hedging framework along with the no-transaction-band network architecture; the accompanying methodology was published in the \textit{Journal of Financial Data Science}. \citet{Francois2024} applied deep hedging to long-term financial derivatives, demonstrating the effectiveness of non-quadratic global hedging policies for risk management of variable annuity guarantees. A comprehensive survey of neural networks for option pricing and hedging is given by \citet{Ruf2020}.

\paragraph{Neural hedging on real data.} \citet{Ruf2021} applied a network they call HedgeNet to end-of-day S\&P~500 options and tick-level Euro Stoxx~50 data, finding that the neural network significantly reduces the mean squared hedging error relative to Black--Scholes delta. However, they also demonstrated that much of this improvement can be replicated by simple linear regressions on option Greeks (delta, vega, vanna), suggesting that the neural network's edge on real data stems primarily from its ability to capture the leverage effect. On simulated Heston data, the improvement was more modest---a finding consistent with our own results and one that motivates the need for uncertainty quantification: if the neural hedger's advantage is sometimes marginal, knowing \textit{when} it is and is not adding value becomes critical.

\paragraph{Uncertainty quantification in deep learning.} Two dominant approaches have emerged for estimating predictive uncertainty in neural networks. MC dropout \citep{Gal2016} interprets dropout at test time as approximate Bayesian inference, producing uncertainty estimates by aggregating multiple stochastic forward passes. Deep ensembles \citep{Lakshminarayanan2017} take a simpler route: multiple networks are trained independently, and their disagreement serves as a proxy for model uncertainty. Empirical comparisons have generally favoured ensembles in terms of calibration quality and out-of-distribution detection, though at the cost of training multiple models. For recurrent architectures, \citet{Gal2016} proposed variational recurrent dropout with a shared mask across time steps, which preserves temporal dependencies that standard dropout disrupts.

\paragraph{Uncertainty quantification in finance.} Bayesian and uncertainty-aware methods have been applied to several problems in quantitative finance, though not, to our knowledge, to the hedging action itself. \citet{Cuchiero2024} used generative adversarial networks for calibration of local stochastic volatility models, producing model parameters consistent with observed market data. Their work addresses uncertainty in the \textit{pricing} problem rather than in the \textit{hedging} problem of choosing a trading strategy. The present paper fills this gap by connecting uncertainty quantification directly to the hedging decision: the ensemble's confidence determines not what the model parameters might be, but how much to trust the model's recommended trading action.

\paragraph{Classical hedging under transaction costs.} The challenge of hedging under transaction costs has a rich analytical literature. \citet{Leland1985} proposed an adjusted delta that accounts for discrete rebalancing and proportional costs by inflating the volatility used in the Black--Scholes formula. \citet{Whalley1997} and \citet{Davis1993} derived the asymptotically optimal no-transaction-band strategy for European options under exponential utility, showing that the optimal policy is to maintain the hedge within a band centred on the Black--Scholes delta and to trade only when the position exits this band. This strategy is theoretically elegant but, as we demonstrate, fragile under model misspecification: when the underlying dynamics depart from the constant-volatility assumption, the band calibration degrades and the strategy can underperform even the naive Black--Scholes delta on tail-risk measures.
\section{Methodology}\label{sec:methodology}

\subsection{Market Model and Problem Setup}\label{sec:setup}

We consider a financial market in which a single risky asset follows the Heston stochastic volatility model \citep{Heston1993}. The spot price $S_t$ and its instantaneous variance $v_t$ evolve according to
\begin{align}
    dS_t &= S_t \sqrt{v_t}\, dW^S_t, \label{eq:heston_spot}\\
    dv_t &= \kappa(\theta - v_t)\,dt + \sigma\sqrt{v_t}\, dW^v_t, \label{eq:heston_var}
\end{align}
under the risk-neutral measure, where $\kappa > 0$ is the rate of mean reversion, $\theta > 0$ is the long-run variance, $\sigma > 0$ is the volatility of variance (vol-of-vol), and the two Brownian motions satisfy $d\langle W^S, W^v \rangle_t = \rho\,dt$ with correlation $\rho \in [-1,1]$. The initial conditions are $S_0 = 1$ and $v_0 = \theta$. We set the risk-free rate $r = 0$, which suppresses the drift term in \eqref{eq:heston_spot} and simplifies the P\&L accounting. The continuous-time dynamics are discretised on an equally spaced grid of $n = 126$ time steps over the trading horizon, and all hedging decisions are taken at these discrete dates.

A trader sells an at-the-money European call option with strike $K = S_0 = 1$ and maturity $T = 0.5$ years (approximately six months, chosen to provide sufficient time for the stochastic volatility dynamics to manifest), and seeks to hedge the resulting exposure by trading in the underlying asset. Each trade incurs a proportional transaction cost at rate $c > 0$: buying or selling $|\Delta \delta_t|$ units at time $t$ costs $c\,S_t\,|\Delta\delta_t|$, where $\Delta \delta_t = \delta_t - \delta_{t-1}$ denotes the change in position. We set $c = 5 \times 10^{-4}$, corresponding to 5 basis points of the traded notional---a level broadly representative of institutional equity markets. The hedging problem is to find a trading strategy $\delta = (\delta_t)_{t=0,\ldots,n}$ that minimises the risk of the terminal profit and loss
\begin{equation}\label{eq:pl}
    \text{P\&L}(\delta) = \sum_{t=0}^{n-1} \delta_t (S_{t+1} - S_t) - \sum_{t=0}^{n} c\,S_t\,|\Delta\delta_t| - (S_n - K)^+,
\end{equation}
where the first term captures hedging gains from stock price movements, the second represents cumulative transaction costs (including the cost of unwinding the final position, with the convention that $\delta_{n+1} = 0$), and the third is the option payoff owed to the buyer.

To examine the robustness of our results, we consider three calibrations of the Heston model, summarised in Table~\ref{tab:calibrations}. The baseline calibration uses parameters broadly consistent with values reported in the empirical literature \citep{AitSahalia2007, Forde2009}. The high vol-of-vol calibration ($\sigma = 0.8$) tests a regime in which the stochastic nature of volatility is particularly pronounced, posing a greater challenge for hedging methods that assume constant volatility. The low correlation calibration ($\rho = -0.3$) weakens the leverage effect that drives much of the interaction between stock returns and volatility dynamics. We note that the high vol-of-vol calibration violates the Feller condition $2\kappa\theta > \sigma^2$, permitting the variance process to reach zero; this is common in empirical calibrations and is handled by the simulation scheme through absorption at zero.

\begin{table}[ht]
\centering
\caption{Heston model calibrations. All calibrations share $\kappa = 2$, $\theta = 0.04$, $v_0 = 0.04$, and proportional transaction cost $c = 5 \times 10^{-4}$.}
\label{tab:calibrations}
\begin{tabular}{lcc}
\toprule
Calibration & $\sigma$ (vol-of-vol) & $\rho$ (correlation) \\
\midrule
Baseline             & 0.4 & $-0.7$ \\
High vol-of-vol      & 0.8 & $-0.7$ \\
Low correlation       & 0.4 & $-0.3$ \\
\bottomrule
\end{tabular}
\end{table}

\subsection{Deep Hedging with Recurrent Networks}\label{sec:deep_hedging}

Following \citet{Buehler2019}, we train a neural network to output the hedge ratio $\delta_t$ at each rebalancing date, given a set of observable market features. The network is trained end-to-end by minimising a convex risk measure applied to the terminal P\&L in \eqref{eq:pl}. We adopt the entropic risk measure
\begin{equation}\label{eq:entropic}
    \rho_a(\text{P\&L}) = \frac{1}{a}\log\mathbb{E}\left[\exp(-a \cdot \text{P\&L})\right],
\end{equation}
with risk-aversion parameter $a = 1$ as the training objective, following the default setting in the \texttt{pfhedge} library \citep{Imaki2023} which provides our implementation framework. Prior to conducting our main experiments, we validated the library against known analytical results under geometric Brownian motion with zero transaction costs: the trained network recovered the Black--Scholes delta with a mean absolute error of 0.035, and the network-implied option price matched the analytical price to within $5 \times 10^{-5}$.

We use a long short-term memory (LSTM) architecture \citep{Hochreiter1997} for the hedging network, motivated by the sequential nature of the hedging problem under stochastic volatility. At each time step, the network receives a feature vector $\mathbf{x}_t = (\log(S_t/K),\; \tau_t,\; \sqrt{v_t},\; \delta_{t-1})$, consisting of the log-moneyness, time to expiry $\tau_t = T - t\Delta t$, the current instantaneous volatility, and the previous hedge ratio. The LSTM maintains a hidden state across time steps, allowing it to condition its output on the recent trajectory of volatility rather than only on the current snapshot.

The architecture consists of a two-layer LSTM with 32 hidden units per layer, followed by a fully connected output layer producing a single hedge ratio $\delta_t \in \mathbb{R}$. We also tested a feedforward network (multilayer perceptron) but found that the LSTM yielded meaningfully better tail-risk performance, consistent with the importance of path information in the presence of stochastic volatility and accumulated transaction costs.

Training proceeds over 500 epochs (300 for the additional calibrations in Section~\ref{sec:robustness}), with 20{,}000 Monte Carlo paths freshly simulated at each epoch. The Adam optimiser \citep{Kingma2015} is used with its default learning rate. Training loss typically stabilises within 200 epochs.

\subsection{Deep Ensembles for Uncertainty Quantification}\label{sec:ensembles}

To quantify the hedging model's uncertainty, we train an ensemble of $M = 5$ independent LSTM networks, each with identical architecture but initialised with different random seeds and trained on independently simulated paths. This follows the deep ensemble methodology of \citet{Lakshminarayanan2017}, which has been shown to produce well-calibrated uncertainty estimates across a range of prediction tasks.

At inference time, all five members produce hedge ratios for the same set of simulated paths. At each time step $t$ and for each path $\omega$, we compute the ensemble mean
\begin{equation}\label{eq:ensemble_mean}
    \bar{\delta}_t(\omega) = \frac{1}{M}\sum_{m=1}^{M}\delta_t^{(m)}(\omega),
\end{equation}
and the ensemble standard deviation
\begin{equation}\label{eq:ensemble_std}
    \psi_t(\omega) = \left(\frac{1}{M-1}\sum_{m=1}^{M}\left(\delta_t^{(m)}(\omega) - \bar{\delta}_t(\omega)\right)^2\right)^{1/2},
\end{equation}
which we interpret as a measure of model uncertainty. A large value of $\psi_t$ indicates substantial disagreement among ensemble members regarding the appropriate hedge at that moment. We note that MC dropout \citep{Gal2016} was also considered but proved ineffective in this setting: the \texttt{pfhedge} framework invokes the model one time step at a time, which limits the effect of dropout applied to single-step outputs.\footnote{Dropout rates of 0.1 and 0.3 both produced average uncertainty below $10^{-3}$. The theoretically appropriate remedy---variational recurrent dropout with a shared mask across time steps \citep{Gal2016}---would require modifications to the library's inference loop and is left for future work.}

\subsection{Classical Baselines}\label{sec:baselines}

We compare the ensemble-based strategy against two classical hedging approaches.

\paragraph{Black--Scholes delta with fixed volatility.} At each time step, the hedge ratio is set to $\delta_t^{\text{BS}} = N(d_1)$, where
\begin{equation}\label{eq:bs_delta}
    d_1 = \frac{\log(S_t/K) + \tfrac{1}{2}\bar{\sigma}^2 \tau_t}{\bar{\sigma}\sqrt{\tau_t}},
\end{equation}
$N(\cdot)$ is the standard normal cumulative distribution function, and $\bar{\sigma} = \sqrt{\theta} = 0.2$ is a fixed volatility estimate equal to the square root of the long-run Heston variance. This represents the strategy of a practitioner who applies the Black--Scholes formula without adapting to the realised volatility path. We verified that replacing $\bar{\sigma}$ with the true instantaneous Heston volatility $\sqrt{v_t}$ has negligible effect on hedging performance (see Section~\ref{sec:true_vol}), indicating that the hedging error under Heston arises not from using an imprecise volatility estimate but from the Black--Scholes model's inability to account for the stochastic evolution of volatility and its correlation with the underlying price.

\paragraph{Whalley--Wilmott no-transaction-band strategy.} The Whalley--Wilmott strategy \citep{Whalley1997, Davis1993} is an asymptotically optimal hedging rule for European options under proportional transaction costs and exponential utility. It defines a no-trade band of half-width
\begin{equation}\label{eq:ww_band}
    w_t = \left(\frac{3\,c\,\Gamma_t^2\,S_t}{2\,a}\right)^{1/3}
\end{equation}
centred on the Black--Scholes delta, where $\Gamma_t$ is the Black--Scholes gamma and $a = 1$ is the risk-aversion parameter. The trader holds the current position if it lies within the band, and moves to the nearest band edge otherwise. While theoretically optimal under Black--Scholes dynamics, the band width is calibrated using the Black--Scholes gamma, which is misspecified under stochastic volatility.

\subsection{Uncertainty-Aware Blending}\label{sec:blending}

The central contribution of this paper is a strategy that blends the ensemble's hedge with the classical Black--Scholes delta, guided by the ensemble's uncertainty. At each time step $t$ and for each path, the blended hedge ratio is
\begin{equation}\label{eq:blend}
    \delta_t^{\text{blend}} = (1 - \alpha_t)\,\bar{\delta}_t + \alpha_t\,\delta_t^{\text{BS}},
\end{equation}
where $\bar{\delta}_t$ is the ensemble mean from \eqref{eq:ensemble_mean}, $\delta_t^{\text{BS}}$ is the Black--Scholes delta from \eqref{eq:bs_delta}, and $\alpha_t \in (0,1)$ is a blending weight parameterised as
\begin{equation}\label{eq:alpha}
    \alpha_t = \text{sigmoid}(\beta_0 + \beta_1\,\psi_t),
\end{equation}
with $\psi_t$ the ensemble uncertainty from \eqref{eq:ensemble_std} and $\beta_0, \beta_1 \in \mathbb{R}$ learnable parameters. A positive $\beta_1$ implies that higher uncertainty shifts the blend toward the classical hedge.

The parameters are optimised on a training set of 7{,}000 paths, with 3{,}000 held out for evaluation. All hedge ratios and uncertainty values are precomputed and held fixed---only $\beta_0$ and $\beta_1$ are updated. We consider two objectives:

\begin{enumerate}
    \item \textit{Entropic risk measure}: minimise $\rho_a(\text{P\&L}^{\text{blend}})$ from \eqref{eq:entropic} with $a = 1$, consistent with the hedger training objective.
    \item \textit{Conditional Value-at-Risk}: minimise $\text{CVaR}_{5\%}(\text{P\&L}^{\text{blend}})$, defined as the expected P\&L conditional on falling in the worst 5\% of outcomes. This directly targets tail risk, the dimension on which the pure ensemble most underperforms classical hedging.
\end{enumerate}

Optimisation uses Adam with a learning rate of 0.01 over 2{,}000 iterations. The blended P\&L is computed in a fully differentiable manner, allowing gradients to flow back to $\beta_0$ and $\beta_1$. We also investigated a richer specification $\alpha_t = \text{sigmoid}(\beta_0 + \beta_1\psi_t + \beta_2 m_t + \beta_3 \tau_t)$ incorporating moneyness $m_t = S_t/K$ and normalised time to expiry; while this produces economically interpretable coefficients (Section~\ref{sec:robustness}), it does not yield a meaningful improvement over the two-parameter model.

\subsection{Evaluation Methodology}\label{sec:evaluation}

All strategies are evaluated on held-out test paths. We report three metrics: mean P\&L (reflecting the expected hedging cost), standard deviation of P\&L (measuring outcome variability), and $\text{CVaR}_{5\%}$ (capturing tail risk). Statistical significance is assessed via paired bootstrap \citep{Efron1993} with 5{,}000 resamples: for each resample, $n$ paths are drawn with replacement and the CVaR difference between strategies is computed on the same resampled paths. The pairing ensures that path-specific variation cancels, yielding a sharper test than unpaired comparisons. We report 95\% confidence intervals and declare significance when the interval excludes zero. Stability is further verified across three independent random seeds for path generation.

\section{Results}\label{sec:results}

We present results for the baseline Heston calibration ($\sigma = 0.4$, $\rho = -0.7$), reserving the remaining calibrations for the robustness analysis in Section~\ref{sec:robustness}. The uncertainty analysis in Sections~\ref{sec:uncertainty_predictive} and~\ref{sec:uncertainty_drivers} uses all 10{,}000 test paths. The blending parameters are optimised on 7{,}000 of these paths, with the remaining 3{,}000 reserved for the strategy comparisons in Sections~\ref{sec:blending_results},~\ref{sec:comparison}, and~\ref{sec:significance}. The P\&L decomposition in Section~\ref{sec:decomposition} uses all 10{,}000 paths.

\subsection{Predictive Power of Ensemble Uncertainty}\label{sec:uncertainty_predictive}

We begin by asking the most fundamental question: does the ensemble's uncertainty carry useful information about hedging performance? To answer this, we sort all test paths by their average uncertainty $\bar{\psi}(\omega) = T^{-1}\sum_t \psi_t(\omega)$ and divide them into five quintiles, from Q1 (most confident) to Q5 (most uncertain). For each quintile, we compute the fraction of paths on which the ensemble mean achieves a better P\&L than the Black--Scholes delta baseline. The results, presented in Table~\ref{tab:winrate}, show a clear monotonic pattern.

\begin{table}[ht]
\centering
\caption{Ensemble win rate against Black--Scholes delta by uncertainty quintile. Win rate denotes the fraction of paths on which the ensemble mean achieves a higher P\&L than the BS delta baseline. Mean advantage is the average P\&L difference (ensemble minus BS delta) within each quintile.}
\label{tab:winrate}
\begin{tabular}{lccc}
\toprule
Quintile & Win Rate (\%) & Mean Advantage & Avg Uncertainty \\
\midrule
Q1 (most confident) & 79.8 & $+0.0140$ & 0.018 \\
Q2                  & 77.5 & $+0.0142$ & 0.026 \\
Q3                  & 64.9 & $+0.0087$ & 0.036 \\
Q4                  & 43.7 & $-0.0013$ & 0.048 \\
Q5 (most uncertain) & 19.3 & $-0.0197$ & 0.066 \\
\midrule
Overall             & 57.0 & $+0.0014$ & 0.039 \\
\bottomrule
\end{tabular}
\end{table}

\begin{figure}[ht]
    \centering
    \includegraphics[width=0.65\textwidth, trim=0 0 0 40, clip]{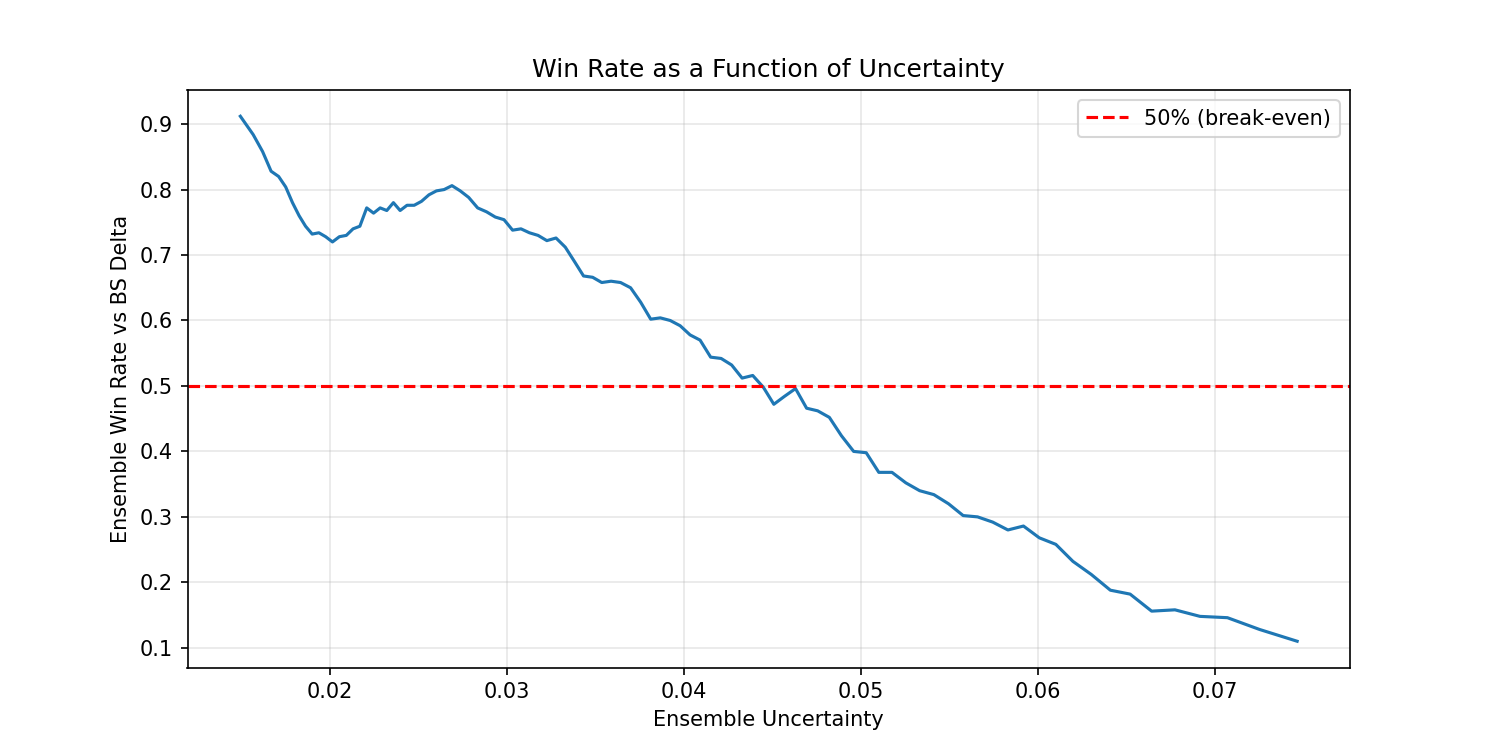}
    \caption{Ensemble win rate against Black--Scholes delta as a function of path-level uncertainty. Each point represents a rolling window of 500 paths sorted by average ensemble standard deviation. The dashed line marks the 50\% break-even threshold.}
    \label{fig:winrate}
\end{figure}

When the ensemble is confident (Q1), it outperforms Black--Scholes delta on approximately 80\% of paths, with an average P\&L advantage of 0.014. When uncertain (Q5), it wins on fewer than 20\% of paths, with an average P\&L disadvantage of $-0.020$. The crossover occurs between Q3 and Q4, at an uncertainty level of roughly 0.04. Figure~\ref{fig:winrate} displays this relationship as a continuous curve, confirming that the decline is smooth and monotonic with no apparent non-linearities.

The correlation between path-level uncertainty and the ensemble's advantage over Black--Scholes delta is $-0.52$, confirming a strong negative association: higher model disagreement reliably predicts worse relative performance. This establishes that the ensemble's uncertainty is not merely noise---it contains actionable information about when the learned strategy should and should not be trusted.

\subsection{Drivers of Ensemble Uncertainty}\label{sec:uncertainty_drivers}

Given that ensemble uncertainty is predictive, it is natural to ask what market conditions give rise to it. Table~\ref{tab:correlations} reports correlations between path-level uncertainty and several market characteristics.

\begin{table}[ht]
\centering
\caption{Correlation between path-level ensemble uncertainty and market characteristics (baseline calibration).}
\label{tab:correlations}
\begin{tabular}{lc}
\toprule
Market Characteristic & Correlation with Uncertainty \\
\midrule
Final moneyness $S_T/K$           & $+0.73$ \\
Average volatility $\overline{\sqrt{v}}$  & $-0.41$ \\
\bottomrule
\end{tabular}
\end{table}

The dominant driver is moneyness, not volatility. Paths on which the stock finishes deep in-the-money (high $S_T/K$) exhibit substantially greater ensemble disagreement than paths ending out-of-the-money. This is the opposite of what one might expect: the ensemble is most uncertain not during volatile market conditions, but during calm, steadily rising markets in which the option moves deep into the money.

We attribute this to the distribution of the training data. Since the option starts at-the-money ($S_0 = K = 1$) and Heston dynamics with negative correlation generate a roughly symmetric distribution of terminal prices, deep in-the-money paths are less frequently encountered during training. The ensemble members, having each seen a somewhat different subset of these rare paths, develop divergent strategies for hedging in this region. By contrast, out-of-the-money paths---where the appropriate hedge ratio is unambiguously close to zero---produce little disagreement regardless of the volatility environment.

\begin{figure}[ht]
    \centering
    \includegraphics[width=0.7\textwidth, trim=0 0 0 48, clip]{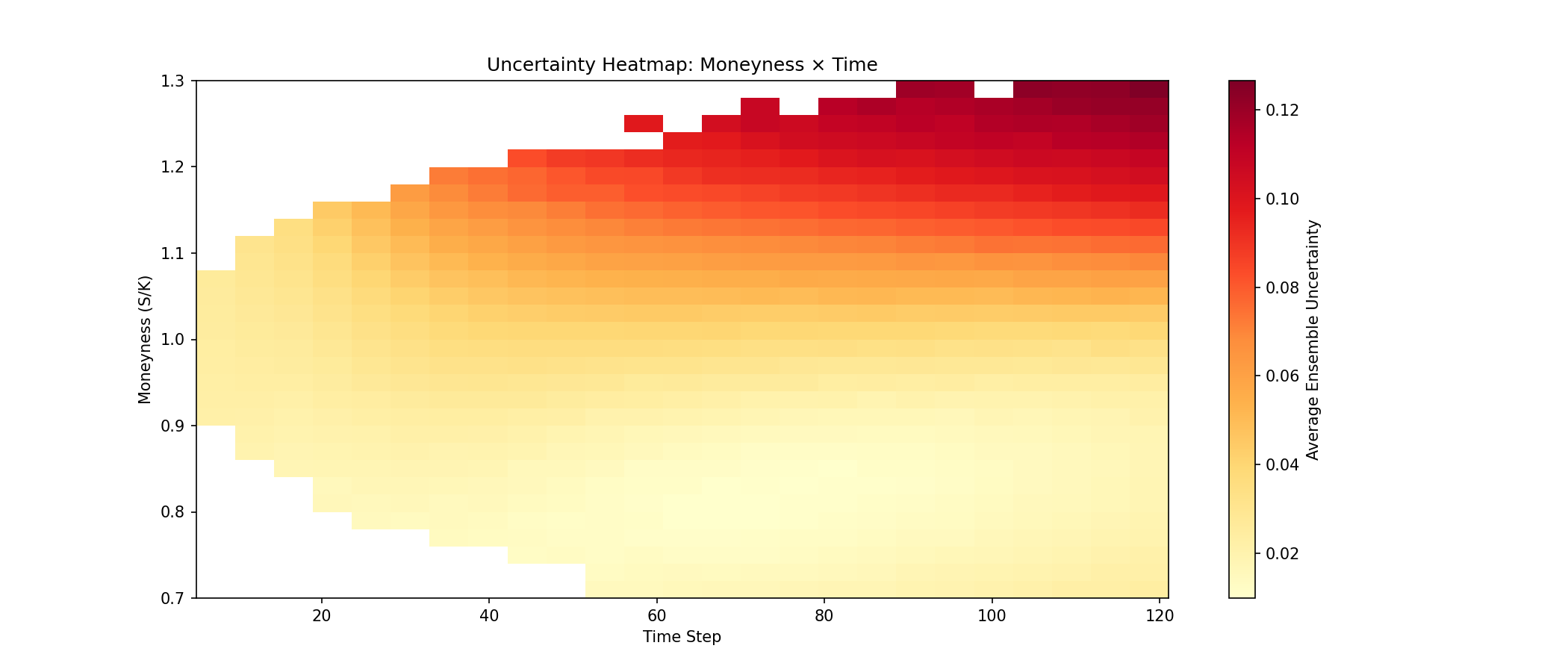}
    \caption{Average ensemble uncertainty as a function of moneyness ($S_t/K$) and time step. Darker colours indicate greater model disagreement. The white region in the lower-left reflects moneyness levels not yet reached by any simulated path at early time steps.}
    \label{fig:heatmap}
\end{figure}

Figure~\ref{fig:heatmap} maps ensemble uncertainty jointly across moneyness and time, revealing a richer picture. Uncertainty is uniformly low early in the option's life (when all paths remain near the money) and grows as paths diverge. The highest uncertainty concentrates in the upper-right region of the heatmap: deep in-the-money paths approaching expiry, precisely where the Black--Scholes delta function is steepest and small differences in learned strategy produce large differences in hedge ratio.

\subsection{Learned Blending Threshold}\label{sec:blending_results}

We optimise the blending parameters $\beta_0$ and $\beta_1$ under two objectives, with markedly different outcomes.

\paragraph{Entropic risk objective.} The optimiser converges to $\beta_0 = -2.24$ and $\beta_1 = -7.56$, yielding $\alpha_t \in [0.04, 0.09]$ across all test paths. The blending weight is close to zero throughout, meaning the strategy almost entirely trusts the ensemble. The negative sign of $\beta_1$ indicates that higher uncertainty pushes the blend further \textit{toward} the ensemble rather than away from it. Under the entropic risk measure, the ensemble's superior mean P\&L outweighs its worse tail risk, and the optimiser sees no benefit in shifting toward the classical hedge.

\paragraph{CVaR objective.} The optimiser converges to $\beta_0 = 0.92$ and $\beta_1 = 0.67$, yielding $\alpha_t \in [0.71, 0.73]$. The strategy now allocates roughly 70\% weight to Black--Scholes delta and 30\% to the ensemble at each time step. The positive $\beta_1$ indicates the expected behaviour: higher uncertainty shifts the blend toward the classical hedge, though the effect is modest given the narrow range of $\alpha_t$.

\begin{figure}[ht]
    \centering
    \includegraphics[width=0.65\textwidth, trim=0 0 0 40, clip]{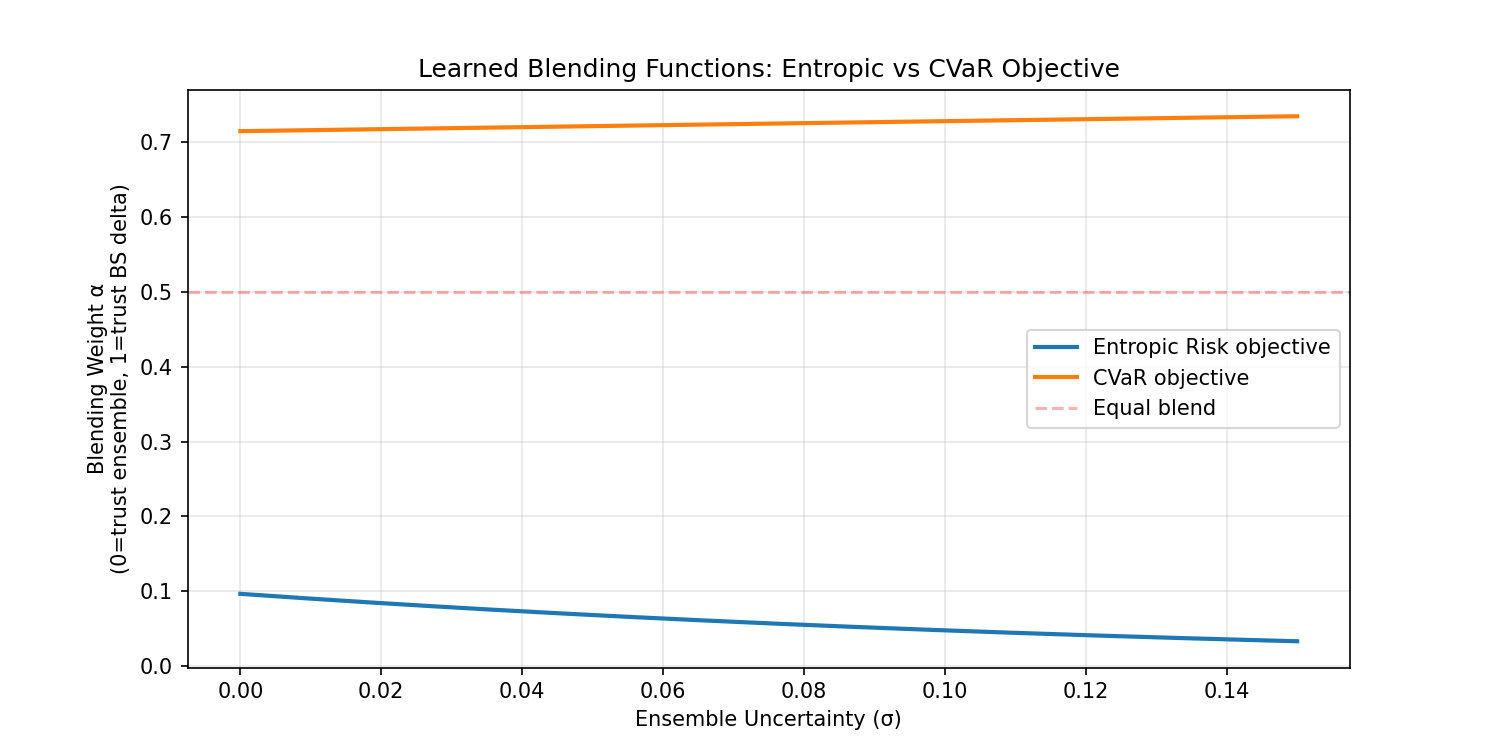}
    \caption{Learned blending functions under two optimisation objectives. The entropic risk objective (lower curve) assigns nearly full weight to the ensemble regardless of uncertainty. The CVaR objective (upper curve) assigns majority weight to Black--Scholes delta, with a slight increase at higher uncertainty levels. The dashed line marks equal weighting.}
    \label{fig:blending_curves}
\end{figure}

The contrast between these two results, displayed in Figure~\ref{fig:blending_curves}, constitutes a result in its own right: the optimal use of uncertainty information depends entirely on the risk manager's objective. The same predictive signal---the same ensemble, the same uncertainty measure---leads to opposite blending decisions depending on whether one prioritises average performance or tail protection. This highlights that uncertainty-aware hedging is not a single strategy but a framework that accommodates different risk preferences.

A notable feature of both learned blending functions is their relative flatness. Rather than a steep sigmoid that switches sharply between strategies at a critical uncertainty threshold, the optimiser consistently finds that a near-constant blend performs best. Under the CVaR objective, the blend allocates approximately 70\% to Black--Scholes delta on every path, with uncertainty contributing only a 2-percentage-point variation. We discuss the economic interpretation of this result in Section~\ref{sec:discussion}.

\subsection{Strategy Comparison}\label{sec:comparison}

Table~\ref{tab:main_results} presents the central comparison across all four strategies under the baseline Heston calibration.

\begin{table}[ht]
\centering
\caption{Hedging performance under the baseline Heston calibration ($\sigma = 0.4$, $\rho = -0.7$). All metrics computed on 3{,}000 held-out test paths. P\&L values are per unit notional.}
\label{tab:main_results}
\begin{tabular}{lccc}
\toprule
Strategy & Mean P\&L & Std P\&L & CVaR$_{5\%}$ \\
\midrule
Pure Ensemble              & $-0.0533$ & $0.0283$ & $-0.1155$ \\
Black--Scholes Delta       & $-0.0551$ & $0.0137$ & $-0.0875$ \\
Whalley--Wilmott           & $-0.0532$ & $0.0208$ & $-0.1037$ \\
CVaR Blend                 & $-0.0545$ & $0.0138$ & $\mathbf{-0.0834}$ \\
\bottomrule
\end{tabular}
\end{table}

Several observations emerge. First, the pure ensemble achieves a strong mean P\&L ($-0.0533$), comparable to Whalley--Wilmott ($-0.0532$), but the worst tail risk (CVaR of $-0.1155$), confirming that the neural hedger adds value on average but is prone to occasional large losses. Second, Black--Scholes delta provides tight risk control (lowest standard deviation, second-best CVaR) at the cost of the worst mean performance. Third, the Whalley--Wilmott strategy---despite being theoretically optimal under Black--Scholes assumptions---performs \textit{worse} than simple Black--Scholes delta on both standard deviation and CVaR under Heston dynamics. Its no-trade band, calibrated to the Black--Scholes gamma, is misspecified under stochastic volatility, leading to suboptimal position adjustments.

The CVaR-optimised blend achieves the best tail risk of any strategy (CVaR of $-0.0834$), improving on Black--Scholes delta by 41 basis points and on Whalley--Wilmott by 203 basis points. It simultaneously achieves a better mean P\&L than Black--Scholes delta ($-0.0545$ vs.\ $-0.0551$), with effectively identical standard deviation. It is the only strategy that dominates Black--Scholes delta on both the mean and CVaR dimensions.

\begin{figure}[ht]
    \centering
    \includegraphics[width=0.75\textwidth, trim=0 0 0 23, clip]{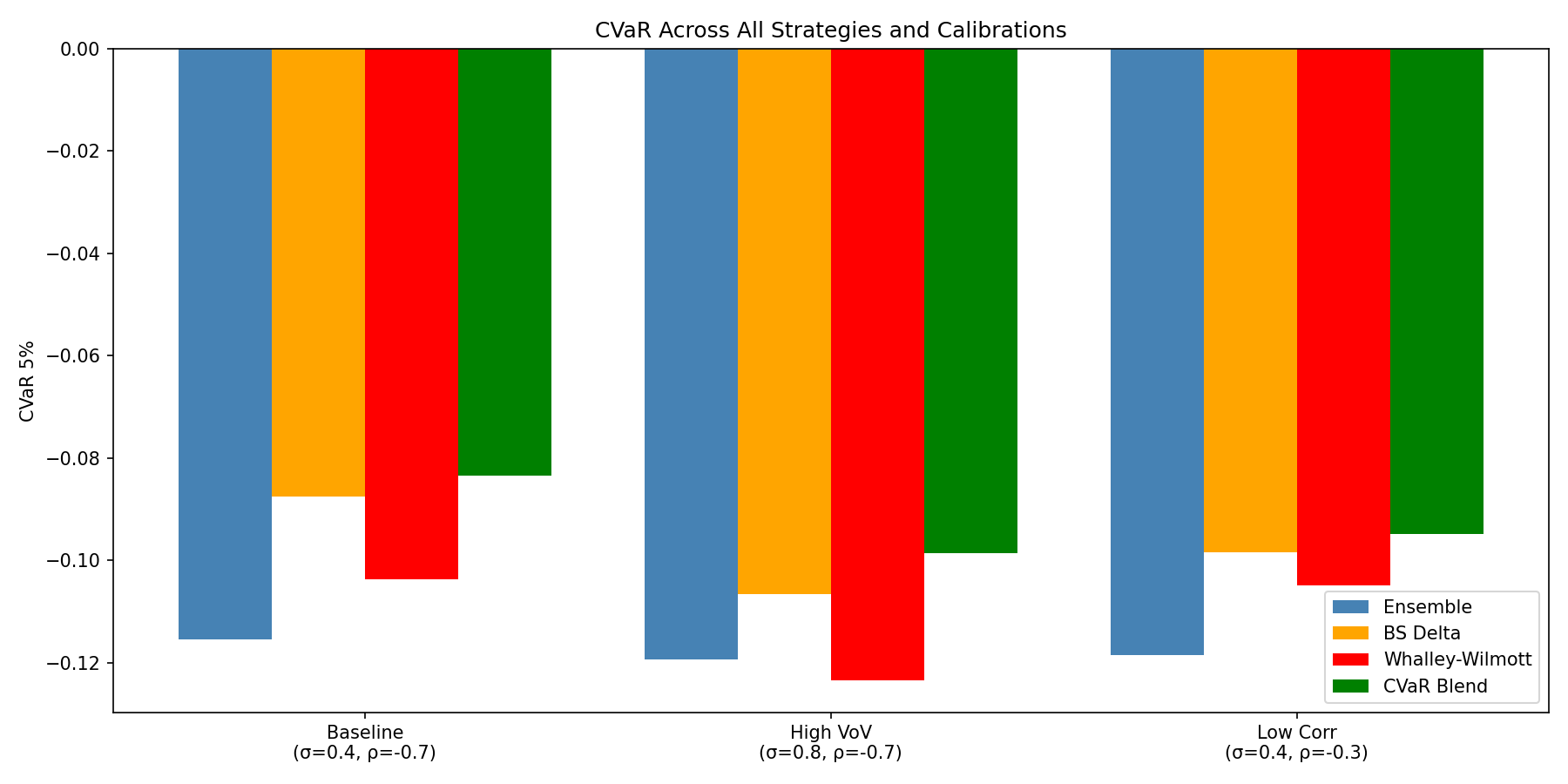}
    \caption{CVaR comparison across all four hedging strategies and all three Heston calibrations. Shorter bars (less negative) indicate better tail-risk performance.}
    \label{fig:pnl_distributions}
\end{figure}

Figure~\ref{fig:pnl_distributions} compares CVaR across all four strategies and all three Heston calibrations. The CVaR blend (green) achieves the least negative CVaR in every calibration. The improvement of the blend over Black--Scholes delta is most pronounced under the high vol-of-vol calibration, where the ensemble contributes the most value.

\subsection{Decomposition of Hedging P\&L}\label{sec:decomposition}

To understand \textit{why} the ensemble achieves a better mean P\&L, we decompose the terminal P\&L into its constituent components: hedging gains from stock price movements, cumulative transaction costs, and the option payoff. Table~\ref{tab:decomposition} reports these components for the ensemble mean strategy and the Black--Scholes delta, computed on the full set of 10{,}000 test paths to improve the precision of the component estimates.

\begin{table}[ht]
\centering
\caption{Decomposition of average P\&L into hedging gains, transaction costs, and option payoff (10{,}000 paths).}
\label{tab:decomposition}
\begin{tabular}{lccc}
\toprule
Component & Ensemble & BS Delta & Difference \\
\midrule
Hedging gains        & $0.0001$ & $0.0003$ & $-0.0002$ \\
Transaction costs    & $0.0006$ & $0.0022$ & $-0.0016$ \\
Option payoff        & $0.0533$ & $0.0533$ & $0.0000$ \\
\midrule
Net P\&L             & $-0.0538$ & $-0.0552$ & $+0.0014$ \\
\bottomrule
\end{tabular}
\end{table}

The ensemble's 14-basis-point advantage arises almost entirely from lower transaction costs ($0.0006$ vs.\ $0.0022$), not from superior hedging gains. In fact, the ensemble's hedging gains are slightly \textit{worse} than those of Black--Scholes delta, indicating that it tracks the option payoff less precisely. The source of its edge is trading efficiency: the ensemble's average trade size per time step is 0.006, compared with 0.026 for Black--Scholes delta---roughly a fourfold reduction in trading activity. The neural network has learned not a better hedging rule, but a more selective one: it identifies moments when adjusting the hedge is not worth the transaction cost, and holds its position instead.

This has direct implications for the blending strategy. The CVaR blend succeeds by combining the tracking accuracy of Black--Scholes delta (which adjusts frequently and follows the true delta closely) with a portion of the ensemble's cost efficiency. At 70\% Black--Scholes delta and 30\% ensemble, the blended strategy trades somewhat less than pure Black--Scholes delta while maintaining most of its tracking precision.

\subsection{Statistical Significance}\label{sec:significance}

The improvements reported above, while consistent and economically meaningful, are modest in absolute terms. We therefore assess their statistical significance using paired bootstrap tests.

\begin{table}[ht]
\centering
\caption{Paired bootstrap tests for CVaR differences (5{,}000 resamples). Positive values indicate the blend has better (less negative) CVaR.}
\label{tab:bootstrap}
\begin{tabular}{lccc}
\toprule
Comparison & Mean Diff (bps) & 95\% CI (bps) & Blend Wins (\%) \\
\midrule
CVaR Blend vs.\ BS Delta        & $+42$ & $[+28,\; +56]$ & 100.0 \\
CVaR Blend vs.\ Whalley--Wilmott & $+181$ & $[+166,\; +195]$ & 100.0 \\
\bottomrule
\end{tabular}
\end{table}

Table~\ref{tab:bootstrap} reports the results. The CVaR improvement of the blend over Black--Scholes delta is 42 basis points with a 95\% confidence interval of $[28, 56]$ basis points, entirely above zero. In all 5{,}000 bootstrap resamples, the blend achieved a better CVaR than Black--Scholes delta. The improvement over Whalley--Wilmott is substantially larger at 181 basis points, also significant at the 95\% level. A separate paired bootstrap test (10{,}000 resamples) for mean P\&L yields a difference of 5 basis points with a 95\% confidence interval of $[3, 6]$, again entirely above zero.

The consistency of these results across bootstrap resamples, combined with stability across three independent random seeds for path generation (Section~\ref{sec:robustness}), establishes that the blend's advantage over classical hedging is not an artefact of a particular sample.
\section{Robustness}\label{sec:robustness}

\subsection{Cross-Calibration Results}\label{sec:cross_calibration}

To assess whether the blending strategy's advantage persists beyond a single parameter configuration, we repeat the full pipeline---training five ensemble members, computing uncertainty, optimising the CVaR blend, and evaluating on held-out paths---for each of the three Heston calibrations described in Table~\ref{tab:calibrations}. Table~\ref{tab:cross_calibration} summarises the results.

\begin{table}[ht]
\centering
\caption{CVaR blend performance across three Heston calibrations. Improvement is measured in basis points relative to Black--Scholes delta. Statistical significance is assessed via paired bootstrap at the 95\% level.}
\label{tab:cross_calibration}
\begin{tabular}{lcccccc}
\toprule
Calibration & Blend CVaR & BS CVaR & Improv.\ (bps) & Signif.\ & Avg $\alpha$ & WW CVaR \\
\midrule
Baseline ($\sigma\!=\!0.4,\,\rho\!=\!{-}0.7$)    & $\mathbf{-0.0834}$ & $-0.0875$ & $+41.5$ & Yes & 0.72 & $-0.1037$ \\
High VoV ($\sigma\!=\!0.8,\,\rho\!=\!{-}0.7$)     & $\mathbf{-0.0987}$ & $-0.1067$ & $+80.1$ & Yes & 0.53 & $-0.1235$ \\
Low Corr ($\sigma\!=\!0.4,\,\rho\!=\!{-}0.3$)     & $\mathbf{-0.0949}$ & $-0.0984$ & $+35.3$ & Yes & 0.65 & $-0.1049$ \\
\bottomrule
\end{tabular}
\end{table}

The blend achieves a statistically significant CVaR improvement over Black--Scholes delta in all three calibrations. The improvement ranges from 35 basis points under low correlation to 80 basis points under high vol-of-vol, with the baseline falling between the two. The blend also substantially outperforms Whalley--Wilmott in every case, with margins of 100 to 250 basis points.

Two patterns merit attention. First, the improvement scales with the severity of model misspecification: the high vol-of-vol calibration, in which the constant-volatility assumption underlying classical hedging is most severely violated, produces the largest gain. The optimiser responds by lowering the average blending weight to $\alpha = 0.53$, allocating nearly half the weight to the ensemble---considerably more than the 72\% Black--Scholes allocation in the baseline. This indicates that the framework automatically adapts the blend to exploit the ensemble more aggressively when the classical model is less reliable.

Second, the Whalley--Wilmott strategy underperforms Black--Scholes delta on CVaR in every calibration, despite being theoretically optimal under constant-volatility assumptions. Under the high vol-of-vol calibration, Whalley--Wilmott achieves a CVaR of $-0.1235$, the worst of any strategy tested. The no-trade band, calibrated to the Black--Scholes gamma, becomes increasingly misspecified as stochastic volatility intensifies.

\subsection{Regime-Dependent Uncertainty}\label{sec:regime_inversion}

The most striking result from the cross-calibration analysis is a qualitative inversion in the relationship between uncertainty and performance under the low-correlation calibration. Table~\ref{tab:winrate_regimes} compares win rates across the three calibrations.

\begin{table}[ht]
\centering
\caption{Ensemble win rate in the most confident (Q1) and most uncertain (Q5) quintiles across calibrations.}
\label{tab:winrate_regimes}
\begin{tabular}{lcc}
\toprule
Calibration & Q1 Win Rate (\%) & Q5 Win Rate (\%) \\
\midrule
Baseline ($\rho = -0.7$)       & 80.7 & 14.4 \\
High VoV ($\rho = -0.7$)       & 80.8 & 29.8 \\
Low Corr ($\rho = -0.3$)       & 12.8 & 89.1 \\
\bottomrule
\end{tabular}
\end{table}

Under strong leverage ($\rho = -0.7$), confidence is a reliable indicator of outperformance: the ensemble wins roughly 80\% of the time on its most confident paths and fewer than 30\% on its least confident. Under weak leverage ($\rho = -0.3$), this relationship reverses entirely---the ensemble wins on nearly 90\% of its most \textit{uncertain} paths and fewer than 13\% of its most confident ones.

The uncertainty drivers invert correspondingly. Table~\ref{tab:uncertainty_drivers} shows that the strong positive correlation between uncertainty and moneyness observed in the baseline ($+0.73$) becomes negative under low correlation ($-0.44$), while the negative correlation with volatility ($-0.41$) reverses to a positive one ($+0.27$).

\begin{table}[ht]
\centering
\caption{Correlation between path-level ensemble uncertainty and market characteristics across calibrations.}
\label{tab:uncertainty_drivers}
\begin{tabular}{lcc}
\toprule
Calibration & Unc.\ vs.\ Moneyness & Unc.\ vs.\ Volatility \\
\midrule
Baseline ($\sigma\!=\!0.4,\,\rho\!=\!{-}0.7$)   & $+0.73$ & $-0.41$ \\
High VoV ($\sigma\!=\!0.8,\,\rho\!=\!{-}0.7$)    & $+0.65$ & $-0.17$ \\
Low Corr ($\sigma\!=\!0.4,\,\rho\!=\!{-}0.3$)    & $-0.44$ & $+0.27$ \\
\bottomrule
\end{tabular}
\end{table}

We interpret this as follows. Under strong leverage, the ensemble learns to exploit the negative stock--volatility correlation, and its confidence reflects how reliably it can do so. High uncertainty on in-the-money paths signals that the ensemble is extrapolating beyond its training distribution---a genuine warning sign. Under weak leverage, there is less systematic structure to exploit. The ensemble's confident predictions in this regime are largely based on surface-level patterns (e.g.\ out-of-the-money options requiring near-zero hedging), while its uncertain predictions reflect genuine diversity in learned strategies that, when averaged, happen to produce a better hedge than any individual member.

The learned blending parameter $\beta_1$ captures this inversion automatically. Under the baseline calibration, $\beta_1 = +0.67$ (higher uncertainty shifts toward Black--Scholes delta). Under low correlation, $\beta_1 = -0.67$ (higher uncertainty shifts \textit{toward} the ensemble). The framework does not require the practitioner to specify the direction of the uncertainty--performance relationship; it discovers it from the data.

\subsection{Stability Across Random Seeds}\label{sec:seed_stability}

To verify that results are not sensitive to the particular random seed used for test path generation, we evaluate the CVaR blend and Black--Scholes delta across three independent seeds. Table~\ref{tab:seed_stability} reports the results for the baseline calibration.

\begin{table}[ht]
\centering
\caption{CVaR blend vs.\ Black--Scholes delta across three random seeds (baseline calibration, 10{,}000 paths each).}
\label{tab:seed_stability}
\begin{tabular}{lccc}
\toprule
Seed & Blend CVaR & BS Delta CVaR & Improvement (bps) \\
\midrule
99  & $-0.0848$ & $-0.0901$ & $+53$ \\
123 & $-0.0852$ & $-0.0907$ & $+55$ \\
456 & $-0.0854$ & $-0.0904$ & $+50$ \\
\bottomrule
\end{tabular}
\end{table}

The improvement is stable across all seeds, with negligible variation. The blend outperforms in every case. The slightly larger magnitude relative to Table~\ref{tab:cross_calibration} reflects the use of the full 10{,}000-path evaluation set rather than the 3{,}000-path held-out subset.

\subsection{Fixed Versus True Volatility in the Classical Baseline}\label{sec:true_vol}

Our primary Black--Scholes baseline uses a fixed volatility $\bar{\sigma} = 0.2$, representing a practitioner who does not observe the Heston instantaneous volatility. One might ask whether giving the classical baseline access to the true volatility $\sqrt{v_t}$ would close the gap. We tested this under the high vol-of-vol calibration ($\sigma = 0.8$), where the discrepancy between fixed and true volatility is largest. Table~\ref{tab:true_vol} shows the results.

\begin{table}[ht]
\centering
\caption{Effect of volatility input on Black–Scholes delta performance and blending (high vol-of-vol calibration, independently simulated paths; minor differences from Table~\ref{tab:cross_calibration} reflect sampling variation).}
\label{tab:true_vol}
\begin{tabular}{lcc}
\toprule
Strategy & Std P\&L & CVaR$_{5\%}$ \\
\midrule
BS Delta (fixed $\bar{\sigma} = 0.2$)   & $0.0239$ & $-0.1125$ \\
BS Delta (true $\sqrt{v_t}$)             & $0.0243$ & $-0.1123$ \\
CVaR Blend (with fixed vol)               & $0.0219$ & $-0.0987$ \\
CVaR Blend (with true vol)                & $0.0228$ & $-0.1039$ \\
\bottomrule
\end{tabular}
\end{table}

Replacing the fixed volatility with the true Heston volatility has a negligible effect on the standalone Black--Scholes delta (CVaR moves from $-0.1125$ to $-0.1123$). The dominant source of hedging error is the structural assumption of constant volatility, not the particular volatility value used. More interesting is the effect on blending: the CVaR blend with fixed-volatility Black--Scholes delta ($-0.0987$) substantially outperforms the blend with true-volatility Black--Scholes delta ($-0.1039$). We attribute this to a diversification effect. The fixed-volatility baseline produces hedge ratios that are more distinct from the ensemble's, providing greater diversification when combined. When the classical and learned strategies respond similarly to current market conditions---as occurs when both use the true volatility---their blend offers less benefit.

\section{Discussion}\label{sec:discussion}

\subsection{The Constant-Mix Finding}

Perhaps the most surprising outcome of the blending optimisation is the flatness of the learned blending function. Under the CVaR objective, the optimiser consistently converges to a near-constant allocation of roughly 70\% Black--Scholes delta and 30\% ensemble, with uncertainty contributing only a 2-percentage-point variation across the full range of observed disagreement levels. We tested several alternative specifications---a richer sigmoid with moneyness and time-to-expiry as additional inputs, and a path-level switching rule with a learned uncertainty threshold---and all converged to approximately the same blend ratio.

This does not mean the uncertainty signal is useless. To the contrary, the quintile analysis in Section~\ref{sec:uncertainty_predictive} demonstrates that it is strongly predictive. The resolution of this apparent paradox lies in the asymmetry of the ensemble's errors. On confident paths (Q1), the ensemble wins on 80\% of occasions, but the remaining 20\% incur losses large enough to substantially worsen the CVaR. A strategy that fully trusts the ensemble on these paths captures the 80\% wins but absorbs the full impact of the 20\% tail losses. The CVaR objective, which averages over the worst 5\% of outcomes, is disproportionately sensitive to these rare but severe losses. The constant 70/30 mix dilutes both the ensemble's wins and its losses, producing a more favourable tail profile than any strategy that concentrates ensemble exposure on nominally confident paths.

The practical implication is that the uncertainty signal's primary value, in the context of CVaR-optimised hedging, lies in \textit{calibrating the optimal blend level} rather than in dynamically switching between strategies. Without the ensemble and its uncertainty measure, a practitioner would have no principled basis for departing from pure Black--Scholes delta. The ensemble provides both the alternative hedge ratios to blend with and the uncertainty measure that, through the optimisation procedure, determines how much weight to place on each.

\subsection{Whalley--Wilmott Under Model Misspecification}

The consistent underperformance of the Whalley--Wilmott strategy relative to Black--Scholes delta on tail risk merits comment. The Whalley--Wilmott no-trade band is derived under the assumption of constant volatility and is calibrated using the Black--Scholes gamma. Under Heston dynamics, this gamma is misspecified: it does not account for the stochastic evolution of volatility or its correlation with the underlying price. As a result, the band width is systematically miscalibrated---too wide in some states and too narrow in others---leading to hedging errors that compound over the option's life.

This observation is consistent with the broader literature on the fragility of asymptotically optimal strategies under model misspecification. It also underscores an advantage of the data-driven approach: the LSTM ensemble, trained directly on Heston-simulated paths, adapts its trading behaviour to the actual dynamics of the market rather than relying on analytical formulae derived under simplified assumptions. The CVaR blend inherits this adaptability while moderating the ensemble's tail risk through the classical component.

\subsection{The Value of the Ensemble}

Table~\ref{tab:individual_members} reports the performance of each individual ensemble member under the baseline calibration. Every member achieves a better mean P\&L than the ensemble mean strategy ($-0.0538$), despite having similar or worse tail risk.

\begin{table}[ht]
\centering
\caption{Individual ensemble member performance (baseline calibration, 10{,}000 paths).}
\label{tab:individual_members}
\begin{tabular}{lccc}
\toprule
Member & Mean P\&L & Std P\&L & CVaR$_{5\%}$ \\
\midrule
Model 1 & $-0.0519$ & $0.0254$ & $-0.1071$ \\
Model 2 & $-0.0521$ & $0.0239$ & $-0.1001$ \\
Model 3 & $-0.0516$ & $0.0227$ & $-0.1011$ \\
Model 4 & $-0.0526$ & $0.0295$ & $-0.1238$ \\
Model 5 & $-0.0517$ & $0.0256$ & $-0.1057$ \\
\midrule
Ensemble Mean & $-0.0538$ & $0.0285$ & $-0.1157$ \\
\bottomrule
\end{tabular}
\end{table}

This is not a contradiction: averaging hedge ratios across models at each time step produces a smoother, less decisive strategy that incurs transaction costs on partial adjustments without fully committing to any single model's trading logic. The ensemble mean is a suboptimal hedging strategy in its own right.

The ensemble's value, therefore, lies not in its mean hedge ratio but in its \textit{disagreement}. The standard deviation across members provides a real-time measure of model confidence that, as we have shown, is predictive of relative performance and useful for calibrating a blended strategy. This suggests that practitioners deploying ensemble-based hedging systems should use the ensemble for uncertainty estimation rather than for generating a consensus hedge.

\subsection{Limitations}

Several limitations of this study should be acknowledged. First, all experiments are conducted on simulated data from the Heston model. While the Heston framework captures key features of equity volatility dynamics---mean-reverting stochastic volatility, leverage effects, and fat-tailed returns---it does not incorporate jumps, regime changes, or the complex term structure and skew dynamics present in real option markets. Extension to real market data, following the methodology of \citet{Ruf2021}, is a natural next step.

Second, we consider only European call options on a single underlying asset. The framework is in principle applicable to exotic payoffs and multi-asset portfolios, but the computational cost of training ensemble members scales with the complexity of the hedging problem.

Third, our uncertainty quantification relies exclusively on deep ensembles. While we documented that standard MC dropout is ineffective in this setting due to the sequential inference structure of the \texttt{pfhedge} library, variational recurrent dropout \citep{Gal2016} remains untested and could provide a computationally cheaper alternative.

Fourth, the blending strategy uses Black--Scholes delta with fixed volatility as the classical component. While we showed that this is preferable to using the true Heston volatility from a diversification standpoint, a practitioner would more realistically use implied volatility derived from option market prices. The interaction between implied-volatility-based hedging and ensemble uncertainty remains unexplored.

Finally, the architecture and hyperparameters of the LSTM ensemble (5 members, 32 hidden units, 2 layers) were not subjected to systematic ablation. While the results are robust across calibrations and random seeds, the sensitivity of the blending improvement to ensemble size, network capacity, and training duration has not been characterised.

\section{Conclusion}\label{sec:conclusion}

This paper has introduced uncertainty quantification to the deep hedging framework, using deep ensembles to produce per-time-step confidence measures for neural hedging strategies. We have shown that ensemble uncertainty is a reliable predictor of when a learned hedging strategy will outperform or underperform classical alternatives, with win rates ranging from over 80\% on confident paths to below 20\% on uncertain ones.

Building on this predictive signal, we proposed a CVaR-optimised blending strategy that combines the ensemble's hedge with the Black--Scholes delta, guided by the level of model disagreement. The blend achieves the best tail risk of any strategy tested---improving on Black--Scholes delta by 35 to 80 basis points in CVaR across three Heston calibrations, and on the Whalley--Wilmott strategy by 100 to 250 basis points---while maintaining a competitive mean P\&L. These improvements are statistically significant and robust.

Beyond the headline performance results, the analysis has revealed several findings of broader relevance. Ensemble uncertainty is driven primarily by moneyness rather than volatility, with models disagreeing most on deep in-the-money paths in calm markets. The relationship between uncertainty and hedging performance is regime-dependent, inverting completely when the leverage effect is weakened. The optimal blending behaviour depends critically on the risk manager's objective: an entropic risk criterion leads to near-complete trust in the ensemble, while a CVaR criterion favours a constant mix dominated by the classical hedge. The Whalley--Wilmott strategy, despite its theoretical optimality under Black--Scholes assumptions, underperforms simple delta hedging under stochastic volatility due to gamma misspecification.

Several extensions merit investigation. The most natural would be validation on real option market data, where the ensemble could potentially exploit richer dynamics---leverage, skew, term structure effects---that are absent from the Heston model. Extension to rough volatility models would test the framework under non-Markovian dynamics where the gap between classical and learned hedging is expected to be larger. On the methodological side, variational recurrent dropout offers a promising alternative to deep ensembles that could reduce the computational overhead of training multiple models. Finally, the blending framework could be extended to incorporate multiple hedging instruments, exotic option payoffs, or adaptive risk preferences that evolve with market conditions.
\newpage
\bibliographystyle{plainnat}
\bibliography{references}
\end{document}